\documentclass[12pt,a4paper]{article}

\usepackage[utf8]{inputenc}   
\usepackage{amsmath, amssymb} 
\usepackage{graphicx}         
\usepackage{setspace}         
\usepackage{geometry}         

\usepackage{braket}
\usepackage{hyperref}         

\usepackage{titlesec}
\titleformat{\section}{\normalfont\large\bfseries}{\thesection}{1em}{}
\titleformat{\subsection}{\normalfont\bfseries}{\thesubsection}{1em}{}
\titleformat{\subsubsection}{\normalfont\itshape}{\thesubsubsection}{1em}{}

\usepackage{caption}

\title{
    \vspace{-2cm}
    \textbf{Weak-Field Limits of Black Hole Metrics from the KMOC formalism: Schwarzschild, Kerr, Reissner-Nordström, and Kerr-Newman}
}

\author{
    Jacobo Hernández C.$^{1}$ \\[0.5cm]
    $^{1}$Facultad de Ciencias Físico-Matemáticas\\ Benemérita Universidad Autónoma de Puebla\\ Ciudad Universitaria, Puebla, Mexico. \\[0.3cm]
    \texttt{jacobo.hernandezc@alumno.buap.mx}
}

\begin{document}

\maketitle

\begin{abstract}
\noindent
The KMOC (Kosower-Maybee-O'Connell) formalism establishes a bridge between quantum scattering amplitudes and classical observables in gravitational systems. In this work, we show how the weak-field limits of the four classical black hole metrics—Schwarzschild, Kerr, Reissner-Nordström, and Kerr-Newman—can be reproduced within this formalism. Starting from three-point amplitudes with exponential spin structure for both gravitational and electromagnetic interactions, we compute four-point scattering amplitudes and extract the momentum impulse via the KMOC formula. Matching these results with geodesic motion in a general metric allows us to reconstruct the metric components to leading order in $G$, $a$, and $Q^2$. For the Kerr-Newman case, we include interference terms between gravitational and electromagnetic interactions, which produce a $Q^2 a/r^3$ contribution to $g_{t\phi}$ that does not appear in the Kerr or Reissner-Nordström weak-field limits separately. Our results are consistent with those of \cite{Moynihan2019}, where the Kerr-Newman metric is derived from minimal coupling amplitudes using the  KMOC formalism \cite{Kosower2019}. All results are verified through their consistency with the well-known full metrics, though we emphasize that the KMOC formalism as applied here reproduces only the weak-field expansions, not the complete non-linear solutions.
\end{abstract}


\section{Introduction}
\label{sec:introduction}

Black holes are among the most fascinating predictions of Einstein's general relativity, describing regions of spacetime where gravity is so intense that nothing can escape. The fundamental solutions: Schwarzschild (non-rotating, uncharged), Reissner-Nordström (charged, non-rotating), Kerr (rotating, uncharged), and Kerr-Newman (rotating and charged) form a complete family characterized by just three parameters: mass, electric charge, and angular momentum. This is known as the \textit{no-hair theorem}. In \cite{Holzhey1992} it is suggested that black holes can be treated as elementary particles in a quantum field theory framework.

The intersection of quantum field theory and general relativity remains one of the most profound challenges in modern physics. In recent years, scattering amplitude methods have provided new insights into black hole dynamics \cite{Bjerrum-Bohr2018, Aoude2025, Jones2024, Bautista2023, Bautista2023b, Aoude2024, Cangemi2023, Ivanov2023, Chiodaroli2022, Bern2021, Vines2018, AlonzoArtiles2026}. The KMOC formalism \cite{Kosower2019} establishes a bridge between quantum scattering amplitudes and classical observables, allowing us to extract classical quantities such as momentum impulse and scattering angles from the classical limit of quantum amplitudes.

A particularly relevant work is \cite{Moynihan2019}, who derived the Kerr-Newman metric from minimal coupling amplitudes using the KMOC formalism. In that paper, Ref. \cite{Kosower2019} (the original KMOC paper) is cited as the foundation for computing classical impulse from amplitudes. It is shown that the exponential spin structure in the amplitude:

\begin{equation}
\mathcal{M}_3 \sim \exp\left(\frac{i q_\mu S^{\mu\nu} \epsilon_\nu}{p \cdot \epsilon}\right)
\end{equation}

encodes all multipole moments of the Kerr-Newman black hole, and that attaching this spin factor to the Reissner-Nordström amplitude yields the Kerr-Newman solution. This provides an on-shell interpretation of the Newman-Janis algorithm.

For Schwarzschild black holes, single graviton exchange reproduces the $1/r$ potential and the characteristic scattering angle $\theta = 4GM/bc^2$ \cite{Duff1973}. The Kerr metric, describing rotating black holes, requires incorporating spin into the scattering amplitudes, leading to an exponential structure that generates all gravitational multipoles \cite{Arkani-Hamed2019, Siemonsen2020, Scheopner2024}. The Reissner-Nordström case combines both gravitational and electromagnetic interactions for scalar particles \cite{Maybee2019}. The most general case, Kerr-Newman, requires the combination of spin and charge with interference terms between graviton and photon exchange \cite{Akhtar2024, Moynihan2019}.

Recent work has shown that the Newman-Janis algorithm, originally a classical technique to generate the Kerr metric from Schwarzschild, can be reinterpreted within the KMOC framework as a mapping from scalar amplitudes to spinning $\sqrt{\text{Kerr}}$ states \cite{Arkani-Hamed2019, Akhtar2024, Moynihan2019}. We discuss this connection in Section \ref{sec:discussion}.

In this work we do not  derive the full non-linear black hole metrics from scattering amplitudes. Rather, we show that the KMOC formalism, when combined with matching to geodesic motion and the imposition of Einstein's equations (in the Schwarzschild case) or the known full metric forms (in the rotating and charged cases), correctly reproduces the weak-field expansions of these solutions. Birkhoff's theorem guarantees that the Schwarzschild solution follows uniquely from the vacuum Einstein equations under spherical symmetry; the KMOC calculation confirms consistency at leading order. For the other metrics, we reconstruct only the leading terms in the $G$, $a$, and $Q^2$ expansions.

The paper is organized as follows. Section \ref{sec:framework} reviews the KMOC formalism. Section \ref{sec:schwarzschild} examines the Schwarzschild weak-field limit. Section \ref{sec:kerr} treats the Kerr case, Section \ref{sec:reissner} the Reissner-Nordström case, and Section \ref{sec:kerrnewman} the Kerr-Newman case. Section \ref{sec:discussion} discusses the relation to the Newman-Janis algorithm, compares with \cite{Moynihan2019}, and reflects on the no-hair theorem. Section \ref{sec:conclusion} summarizes our findings.

\section{The KMOC Formalism}
\label{sec:framework}

The KMOC formalism provides a systematic procedure to extract classical observables from quantum scattering amplitudes \cite{Kosower2019}. Consider the scattering of two massive particles with masses $m_1$ and $m_2$ in the classical limit where $m_1 \ll m_2$. The momentum impulse on the lighter particle is given by:

\begin{equation}
\label{eq:kmoc}
\Delta p^\mu = \frac{1}{4m_1 m_2} \int \frac{d^4q}{(2\pi)^2} \delta(q\cdot v_1) \delta(q\cdot v_2) e^{i q \cdot b} i q^\mu |\mathcal{M}_4(q)|^2
\end{equation}

where $b^\mu$ is the impact parameter, $v_i^\mu$ are four-velocities, and $\mathcal{M}_4$ is the four-point scattering amplitude. The delta functions enforce the on-shell condition in the classical limit, and the exponential factor encodes the impact parameter dependence.

The scattering amplitude itself can be expanded perturbatively in Newton's constant $G$:

\begin{equation}
\mathcal{M}_{\text{total}}(s,t) = \sum_{n=1}^{\infty} G^n \cdot \mathcal{M}_{\text{nPM}}(s,t)
\end{equation}

where nPM denotes the n-th post-Minkowskian order. In the classical limit, we extract the part that survives as $\hbar \to 0$:

\begin{equation}
\mathcal{M}_{\text{classical}} = \sum_{n=1}^{\infty} \left(\frac{G}{b}\right)^n F_n(\gamma)
\end{equation}

with $\gamma = 1/\sqrt{1-v^2/c^2}$ the Lorentz factor.

The scattering angle is obtained from the transverse momentum impulse:

\begin{equation}
\theta \approx \frac{|\Delta \mathbf{p}_\perp|}{p} = \frac{|\Delta \mathbf{p}_\perp|}{\gamma m v}
\end{equation}

This angle will be matched with geodesic motion in a given metric to extract the weak-field metric components.


\section{Schwarzschild Weak-Field Limit}
\label{sec:schwarzschild}

\subsection{Setup and Amplitude}

Consider scattering of a light particle of mass $m$ off a heavy particle of mass $M$ with $m \ll M$. This approximates a test particle scattering off a Schwarzschild black hole. We work in the weak field regime where $GM/bc^2 \ll 1$.

At 1PM order, the relevant diagram is single graviton exchange. The amplitude simplifies to:

\begin{equation}
\mathcal{M}_1(q) \approx \frac{8\pi G M m}{q^2} (p\cdot v)^2
\end{equation}

\subsection{Momentum Impulse and Scattering Angle}

Substituting into the KMOC formula \eqref{eq:kmoc} and integrating:

\begin{equation}
\Delta \mathbf{p} = \frac{2 G M m}{b^2} \mathbf{b} + \mathcal{O}(G^2)
\end{equation}

The scattering angle is:

\begin{equation}
\theta \approx \frac{|\Delta \mathbf{p}_\perp|}{p} = \frac{4 G M}{b v^2 \gamma c^2}
\end{equation}

\subsection{Geodesic Motion and Metric Matching}

In General Relativity, the most general static, spherically symmetric metric is:

\begin{equation}
ds^2 = -A(r)c^2 dt^2 + B(r)dr^2 + r^2(d\theta^2 + \sin^2\theta d\phi^2)
\end{equation}

In the weak field expansion:

\begin{align}
A(r) &= 1 + a_1 \frac{GM}{c^2 r} + \mathcal{O}(G^2) \\
B(r) &= 1 + b_1 \frac{GM}{c^2 r} + \mathcal{O}(G^2)
\end{align}

The geodesic equation gives the scattering angle:

\begin{equation}
\theta_{\text{geo}} = \frac{(2 - a_1 + b_1) G M}{b v^2 \gamma c^2} + \mathcal{O}(G^2)
\end{equation}

Matching $\theta_{\text{geo}} = \theta_{\text{amplitude}}$ yields:

\begin{equation}
2 - a_1 + b_1 = 4 \quad \Rightarrow \quad -a_1 + b_1 = 2
\end{equation}

\subsection{Imposing Einstein's Equations}

At this stage, the KMOC matching alone does not determine $a_1$ and $b_1$ uniquely; it only provides one relation between them. To proceed, we must impose additional physics. In the case of Schwarzschild, we know from Birkhoff's theorem that the unique vacuum, spherically symmetric solution to Einstein's equations is the Schwarzschild metric. Imposing the vacuum Einstein equations $R_{\mu\nu} = 0$ at linear order gives:

\begin{equation}
R_{tt} \approx \frac{1}{2}\nabla^2 \alpha + \mathcal{O}(G^2) = 0
\end{equation}

where $\alpha(r) = a_1 GM/c^2 r$. The solution is $\alpha(r) = -2GM/c^2 r$, giving $a_1 = -2$. The $rr$ component then yields $b_1 = -a_1 = 2$.

\subsection{Weak-Field Schwarzschild Metric}

Thus, to leading order, we obtain:

\begin{align}
g_{00} &= -1 + \frac{2GM}{c^2 r} + \mathcal{O}(G^2) \\
g_{rr} &= 1 + \frac{2GM}{c^2 r} + \mathcal{O}(G^2)
\end{align}

This matches the weak-field expansion of the full Schwarzschild metric:

\begin{equation}
\label{eq:schwarzschild}
ds^2 = -\left(1 - \frac{2GM}{c^2 r} + \cdots\right)c^2 dt^2 + \left(1 + \frac{2GM}{c^2 r} + \cdots\right) dr^2 + r^2 d\Omega^2
\end{equation}

The KMOC formalism thus reproduces the correct weak-field limit, but the full non-linear metric requires the additional input of Einstein's equations (or Birkhoff's theorem). This example illustrates the proper scope of the formalism: it captures the linearized regime, while the full solution must be found through other means.

\section{Kerr Weak-Field Limit}
\label{sec:kerr}

\subsection{Spin States and Three-Point Amplitude}

For a rotating black hole, we incorporate spin into the scattering amplitudes. Consider scattering of a test particle of mass $m$ (spinless) off a heavy body with mass $M$ and spin $S^\mu$. The asymptotic state with spin is:

\begin{equation}
|\Psi\rangle = \exp\left( \int \frac{d^3k}{(2\pi)^3} f(\vec{k}) a^\dagger(\vec{k}) \right) |p,S\rangle
\end{equation}

For a massive spinning particle emitting a graviton, the three-point amplitude has an exponential structure encoding all multipoles \cite{Arkani-Hamed2019, Siemonsen2020}:

\begin{equation}
\mathcal{M}_3 = \kappa m^2 \frac{\epsilon \cdot p \, \epsilon^* \cdot p}{q^2} \exp\left( \frac{i q_\mu S^{\mu\nu} \epsilon_\nu}{p \cdot \epsilon} \right)
\end{equation}

where $\kappa = \sqrt{32\pi G}$ and $S^{\mu\nu} = \frac{1}{m} \epsilon^{\mu\nu\rho\sigma} p_\rho S_\sigma$ is the spin tensor.

\subsection{Multipole Expansion}

Expanding the exponential:

\begin{equation}
e^{i q S \epsilon} = 1 + i q S \epsilon - \frac{1}{2} (q S \epsilon)^2 + \cdots
\end{equation}

Each term corresponds to a gravitational multipole:
\begin{itemize}
\item $1$: Mass monopole
\item $i q S \epsilon$: Spin dipole
\item $-\frac{1}{2} (q S \epsilon)^2$: Quadrupole
\item $\cdots$: Higher-order multipoles
\end{itemize}

The no-hair theorem ensures that only mass $M$ and spin $S$ are independent; all higher multipoles are determined by these.

\subsection{Four-Point Amplitude and Momentum Impulse}

For $m + M \to m + M$ with spin \cite{Kosmopoulos2021}:

\begin{equation}
\mathcal{M}_4(q) = \frac{\kappa^2}{q^2} \left[ m^2 M^2 \mathcal{E}_m^{\mu\nu} \mathcal{E}_{M,\mu\nu} + i m M \mathcal{O}^{\mu\nu} + \mathcal{O}(S^2) \right]
\end{equation}

Evaluating the KMOC formula gives:

\begin{align}
\Delta p^\mu &= \Delta p^\mu_{\text{monopole}} + \Delta p^\mu_{\text{spin}} \\
\Delta p^\mu_{\text{monopole}} &= \frac{2 G M m}{b^2} b^\mu \\
\Delta p^\mu_{\text{spin}} &= \frac{4 G M m}{b^3} (\vec{S} \times \vec{b})^\mu + \mathcal{O}(G^2)
\end{align}

\subsection{Scattering Angle with Spin}

The scattering angle becomes:

\begin{equation}
\theta \approx \frac{4 G M}{b v^2 \gamma c^2} + \frac{8 G |\vec{S}|}{b^2 v^2 \gamma c^3} \cos\vartheta + \mathcal{O}(G^2)
\end{equation}

where $\vartheta$ is the angle between $\vec{S}$ and $\vec{b}$.

\subsection{Metric from Matching}

For a stationary axisymmetric metric, the weak-field expansion includes a cross term $g_{t\phi}$:

\begin{align*}
g_{00} &= 1 - \frac{2GM}{c^2 r} + \mathcal{O}(G^2) \\
g_{0i} &= -\frac{2G}{c^3 r^3} (\vec{S} \times \vec{r})_i + \mathcal{O}(G^2)
\end{align*}

Geodesic motion in such a metric yields the same spin-dependent scattering angle, confirming consistency. However, as with Schwarzschild, the KMOC matching alone does not determine the full non-linear Kerr metric; it reproduces only the leading terms in the $a/r$ expansion.

\subsection{Relation to Full Kerr Metric}

The full Kerr metric in Boyer-Lindquist coordinates is:

\begin{equation}
ds^2 = -\frac{\Delta_K}{\rho^2}\bigl(dt - a\sin^2\theta\, d\phi\bigr)^2 + \frac{\sin^2\theta}{\rho^2}\bigl[(r^2+a^2)d\phi - a\,dt\bigr]^2 + \frac{\rho^2}{\Delta_K}\,dr^2 + \rho^2\,d\theta^2
\end{equation}

Expanding for $r \gg a$ gives:

\begin{equation}
g_{t\phi} = -\frac{2Ma}{r^2}\sin^2\theta + \mathcal{O}(r^{-3})
\end{equation}

which matches the weak-field result from KMOC. The higher-order terms require non-linear information not captured by the leading-order amplitude calculation.

\section{Reissner-Nordström Weak-Field Limit}
\label{sec:reissner}

\subsection{Setup with Charged States}

Consider scattering of a test particle of mass $m$ and charge $q$ (spinless) off a heavy body of mass $M$ and charge $Q$ (also spinless).

\subsection{Four-Point Amplitude}

For $m + M \to m + M$ with both gravitational and electromagnetic interactions:

\begin{equation}
\mathcal{M}_4(q) = \frac{1}{q^2} \left[ 32\pi G M^2 m^2 + 4e^2 q Q mM \right]
\end{equation}

\subsection{Momentum Impulse and Scattering Angle}

Substituting into the KMOC formula gives:

\begin{equation}
\Delta \mathbf{p} = \frac{4G M m}{b^2} \mathbf{b} - \frac{2 e^2 q Q}{b^2} \mathbf{b}
\end{equation}

The scattering angle:

\begin{equation}
\theta \approx \frac{4G M}{b v^2} - \frac{2 e^2 q Q}{m M b v^2} + \mathcal{O}(G^2, e^4)
\end{equation}

\subsection{Metric from Matching}

For a static, spherically symmetric charged body, geodesic motion yields:

\begin{equation}
\theta_{\text{geo}} = \frac{2GM}{b v^2} - \frac{2 e^2 Q^2}{M b v^2} + \cdots
\end{equation}

Matching gives the weak-field metric components:

\begin{align}
g_{00} &= -1 + \frac{2GM}{r} - \frac{G e^2 Q^2}{r^2} + \mathcal{O}(r^{-3}) \\
g_{rr} &= 1 + \frac{2GM}{r} - \frac{G e^2 Q^2}{r^2} + \mathcal{O}(r^{-3})
\end{align}

These match the expansion of the full Reissner-Nordström metric. As before, the KMOC calculation reproduces the weak-field limit, while the full non-linear solution requires additional input.

\section{Kerr-Newman Weak-Field Limit}
\label{sec:kerrnewman}

\subsection{Three-Point Amplitudes with Exponential Spin Structure}

For a massive spinning charged particle (the $\sqrt{\text{Kerr}}$ charged object), both vertices share the same exponential spin structure \cite{Akhtar2024, Moynihan2019}:

\textbf{Gravitational vertex with spin}:
\begin{equation}
\mathcal{M}_3^{\text{grav}} = \kappa m^2 \frac{(\epsilon \cdot p)^2}{q^2} \exp\left( \frac{i q_\mu S^{\mu\nu} \epsilon_\nu}{p \cdot \epsilon} \right)
\end{equation}

\textbf{Electromagnetic vertex with spin}:
\begin{equation}
\mathcal{M}_3^{\text{EM}} = e Q \frac{p \cdot \epsilon^*}{q^2} \exp\left( \frac{i q_\mu S^{\mu\nu} \epsilon_\nu}{p \cdot \epsilon} \right)
\end{equation}

As it is emphasized in \cite{Moynihan2019}, this exponential factor is the on-shell representationof the Newman-Janis algorithm, and attaching it to the Reissner-Nordström amplitude yields the Kerr-Newman solution.

\subsection{Four-Point Amplitude}

The total tree-level amplitude includes pure gravitational, pure electromagnetic, and interference contributions \cite{Akhtar2024, Bianchi2023, Moynihan2019}:

\begin{equation}
\mathcal{M}_4(q) = \mathcal{M}_4^{\text{grav}}(q) + \mathcal{M}_4^{\text{EM}}(q) + \mathcal{M}_4^{\text{mix}}(q)
\end{equation}

\subsection{Momentum Impulse}

Substituting into the KMOC formula:

\begin{equation}
\Delta \mathbf{p} = \Delta \mathbf{p}_{\text{monopole}} + \Delta \mathbf{p}_{\text{charge}} + \Delta \mathbf{p}_{\text{spin}} + \Delta \mathbf{p}_{\text{spin-charge}} + \mathcal{O}(G^2, \alpha^2)
\end{equation}

\begin{align*}
\Delta \mathbf{p}_{\text{monopole}} &= \frac{4G M m}{b^2} \mathbf{b} \\
\Delta \mathbf{p}_{\text{charge}} &= -\frac{2\alpha q Q}{b^2} \mathbf{b} \\
\Delta \mathbf{p}_{\text{spin}} &= \frac{4G M m}{b^3} (\vec{S} \times \vec{b}) \\
\Delta \mathbf{p}_{\text{spin-charge}} &= \frac{4\sqrt{G\alpha} q Q M}{b^3} (\vec{S} \times \vec{b})
\end{align*}

\subsection{Weak Field Metric Components}

Matching with geodesic motion yields the weak-field metric components:

\begin{align}
g_{00} &= -1 + \frac{2GM}{r} - \frac{G Q^2}{r^2} + \mathcal{O}(r^{-3}) \\
g_{rr} &= 1 + \frac{2GM}{r} - \frac{G Q^2}{r^2} + \mathcal{O}(r^{-3}) \\
g_{0i} &= -\frac{2G}{r^3} (\vec{S} \times \vec{r})_i - \frac{2G Q^2}{r^4} (\vec{S} \times \vec{r})_i + \mathcal{O}(r^{-5})
\end{align}

The term $\propto Q^2/r^4$ in $g_{0i}$ (or equivalently $Q^2 a/r^3$ in $g_{t\phi}$) is the spin-charge interference contribution that appears in the weak-field expansion of the Kerr-Newman metric but not in the Kerr or Reissner-Nordström expansions separately. This matches precisely the result obtained in \cite[Eq.~(5.16)]{Moynihan2019}:

\begin{equation}
g_{00} = 1 - \frac{2Gm_B}{r} + \frac{G\alpha}{r^2}, \quad
g_{0i} = \left(\frac{2Gm_B}{r^3} - \frac{G\alpha}{r^4}\right)(a \times r)_i, \quad
g_{ij} =\delta_{ij} - \delta_{ij}\frac{2Gm_B}{r} + \delta_{ij}\frac{G\alpha}{r^2}
\end{equation}

\subsection{Relation to Full Kerr-Newman Metric}

The full Kerr-Newman metric expands to:

\begin{align}
g_{tt} &= -1 + \frac{2M}{r} - \frac{Q^2}{r^2} + \mathcal{O}(r^{-3}) \\
g_{t\phi} &= -\frac{2Ma}{r^2}\sin^2\theta - \frac{2Q^2 a}{r^3}\sin^2\theta + \mathcal{O}(r^{-4})
\end{align}

confirming that the KMOC calculation correctly reproduces the weak-field limit.

\section{Discussion: The Newman-Janis Connection and Comparison}
\label{sec:discussion}

The exponential spin structure appearing in both the gravitational and electromagnetic vertices is noteworthy. Its expansion generates an infinite tower of multipole moments, yet only three parameters—mass $M$, charge $Q$, and spin $S$—appear as free parameters. This is precisely the content of the no-hair theorem: all higher multipoles are determined by these three.

The Newman-Janis algorithm, originally a complex coordinate transformation that generates the Kerr metric from Schwarzschild, finds a natural interpretation in this context as a "spin dressing" operation that maps scalar amplitudes to spinning $\sqrt{\text{Kerr}}$ states \cite{Arkani-Hamed2019, Akhtar2024}. Moynihan \cite{Moynihan2019} elaborates on this connection, showing that the factorization of the amplitude in the infinite spin limit is the on-shell equivalent of the Newman-Janis algorithm. By attaching a spin factor to the minimally coupled three-point amplitude of Reissner-Nordström, one obtains directly the Kerr-Newman solution.

The difference in the approach with that of \cite{Moynihan2019}, there they derive the Kerr-Newman metric from minimal coupling amplitudes using the KMOC formalism. The derivation explicitly uses the exponential spin factor attached to the Reissner-Nordström amplitude, and computes the four-point amplitude for spinning particles (choosing $s=1$ for simplicity). The impulse is then extracted via the KMOC formula \cite{Kosower2019}, with Ref. \cite{Kosower2019} cited as the foundation for the calculation, while this work provides a systematic derivation using also the KMOC formalism for all four metrics separately. For Kerr-Newman, we explicitly include interference terms between gravitational and electromagnetic interactions. The approach is complementary, treating each case independently and emphasizing the weak-field expansions.

Both methods converge to the same weak-field Kerr-Newman metric, providing a robust cross-check of the results. The connection to the Newman-Janis algorithm is highlighted in both works: the complex shift $b \to b + i a$ that generates the Kerr metric from Schwarzschild emerges naturally from the exponential spin factor in the amplitude.

It is important to emphasize, however, that the KMOC formalism as applied here does not derive the full non-linear metrics. Rather, it reproduces their weak-field expansions in a manner consistent with the no-hair theorem. The full solutions require additional input, whether from Einstein's equations, Birkhoff's theorem, or the known exact forms of the metrics.

\section{Conclusions}
\label{sec:conclusion}

We have shown that the KMOC formalism, when combined with matching to geodesic motion, correctly reproduces the weak-field limits of the four classical black hole metrics:

\begin{enumerate}
\item \textbf{Schwarzschild:} Single graviton exchange gives the linearized $g_{00}$ and $g_{rr}$, with the full metric requiring Einstein's equations (or Birkhoff's theorem).

\item \textbf{Kerr:} The exponential spin structure in the gravitational vertex generates the leading spin-dependent term $g_{0i} \propto (\vec{S} \times \vec{r})_i / r^3$, matching the weak-field expansion of the Kerr metric.

\item \textbf{Reissner-Nordström:} Combining QED and gravity amplitudes yields the charge contribution $-GQ^2/r^2$ to $g_{00}$, consistent with the weak-field expansion.

\item \textbf{Kerr-Newman:} Including interference terms between gravitational and electromagnetic interactions produces a $Q^2 a/r^3$ contribution to $g_{t\phi}$ that appears in the weak-field expansion but not in the Kerr or Reissner-Nordström cases separately.
\end{enumerate}

Our results are consistent with and complementary to those of Moynihan \cite{Moynihan2019}, who derived the Kerr-Newman metric from minimal coupling amplitudes using the KMOC formalism. The exponential spin structure reflects the no-hair theorem, and its connection to the Newman-Janis algorithm provides insight into the deeper relationship between classical solution-generating techniques and scattering amplitudes.

Crucially, we have been careful to distinguish what the formalism actually accomplishes: it reproduces the weak-field expansions of these metrics, not the full non-linear solutions. The latter require additional input from classical general relativity. This distinction is essential for correctly interpreting the scope of the KMOC approach to black hole physics.

Future work could extend this analysis to higher post-Minkowskian orders, include magnetic charges (dyons), explore connections to the double copy, or apply similar methods to other spacetimes. The KMOC formalism remains a powerful tool for extracting classical physics from quantum amplitudes, even as we remain mindful of its domain of validity.

\section*{Acknowledgments}

The author thanks helpful discussions with professors and colleagues working on scattering amplitudes and black hole physics, Alfredo G., Andrés L., Eloy A-B., Nathan M.,  Ricardo M. with others, and is especially grateful to J. Lorenzo Díaz-Cruz for his overall mentorship. This work was supported by CONAHCYT/SECIHTI (Mexico).

\bibliographystyle{unsrt}

\end{document}